\documentclass[12pt,preprint]{aastex}

\begin{document}

\title{Pulsation and evolutionary masses of classical Cepheids. I. Milky Way variables}

\author{F. Caputo \altaffilmark{1}, G. Bono \altaffilmark{1}, G. Fiorentino \altaffilmark{1,2},
M. Marconi \altaffilmark{3}, I. Musella \altaffilmark{3}}

\affil{1. INAF $-$ Osservatorio Astronomico di Roma, Via Frascati 33, 00040 Monte Porzio Catone, 
Italy; caputo@mporzio.astro.it; bono@mporzio.astro.it}

\affil{2. Universit\`a di Roma Tor Vergata, via della Ricerca Scientifica 1, 00133 Roma, Italy; 
giuliana@mporzio.astro.it}

\affil{3. INAF $-$ Osservatorio Astronomico di Capodimonte, Via Moiariello 16, 80131 Napoli, Italy; 
marcella@na.astro.it; ilaria@na.astro.it}

\begin{abstract}
We investigate a selected sample of Galactic classical Cepheids with
available distance and reddening estimates in the framework of the
theoretical scenario provided by pulsation models, computed with metal 
abundance $Z$=0.02, helium content in the range of $Y$=0.25 to 0.31, 
and various choices of the stellar mass and luminosity. After
transforming the bolometric light curve of the fundamental models into
$BVRIJK$ magnitudes, we derived analytical relations connecting the 
pulsation period with the stellar mass, the mean (intensity-averaged) 
absolute magnitude, and the color of the pulsators. These relations 
are used together with the Cepheid observed absolute magnitudes 
in order to determine the ``pulsation" mass -$M_p$- of each individual
variable. The comparison with the ``evolutionary" masses -$M_{e,can}$-  
given by canonical (no convective core overshooting, no mass-loss) models 
of central He-burning stellar structures reveals that the $M_p/M_{e,can}$
ratio is correlated with the Cepheid period, ranging from $\sim$ 0.8 at
log$P$=0.5 to $\sim$ 1 at log$P$=1.5. We discuss the effects of different
input physics and/or assumptions on the evolutionary computations, as 
well as of uncertainties in the adopted Cepheid metal content, 
distance, and reddening. Eventually, we find that the pulsational 
results can be interpreted in terms of mass-loss during or before 
the Cepheid phase, whose amount increases as the Cepheid original mass 
decreases. It vanishes around 13$M_{\odot}$ and increases up to $\sim$ 20\% at 4$M_{\odot}$.

\keywords{Stars: variables: Cepheids -- Stars:  oscillations -- Stars: mass loss}
\end{abstract}

\pagebreak

\section{Introduction}
Classical Cepheids have long been recognized as primary standard
candles to estimate the distance of external galaxies out to the
Virgo cluster. Moreover, through the calibration of secondary
distance indicators, they allow the investigation of even more
remote stellar systems, thus enabling us to obtain information 
on the Hubble
constant (Ferrarese et al. 2000; Freedman et al. 2001; Saha et al. 2001). 
However, their importance exceeds the determination of distances, since 
they are powerful astrophysical laboratories providing fundamental
clues for studying the evolution of intermediate-mass stars and,
in particular, the occurrence of mass-loss along the Red Giant
(RG) and the central He-burning evolutionary phases.

From the point of view of the stellar evolution theory, Cepheids 
are indeed generally interpreted as post-RG stars crossing the
pulsation region of the HR diagram during the characteristic ``blue
loop" connected with core He-burning. During this phase, the
luminosity $L$ of the evolutionary track mainly depends on the
original stellar mass $M$ and the chemical composition:
therefore, the evolutionary models computed by neglecting
the mass-loss, provide a Mass-Luminosity ($ML$) relation, which is
widely used to estimate the ``evolutionary" mass of Cepheids for
which absolute magnitudes and chemical composition are available.
On the other hand, the Cepheid pulsation period depends, 
at fixed chemical composition, on the star mass, luminosity and
effective temperature, hence the ensuing mass-dependent
Period-Luminosity-Color ($PLC$) relation can be used to estimate
the ``pulsation" mass of each individual variable with known metal
content, absolute magnitude and intrinsic color. With a slightly 
different approach, the theoretical Period-Mass-Radius ($PMR$)
relation can be applied to Cepheids for which accurate estimates
of radii are available.

In the last decades, a large amount of work has been devoted to
the comparison between pulsation and evolutionary masses, leading
to the long-debated problem of the ``Cepheid mass discrepancy'' (see
Cox 1980). Almost all the studies suggest that the pulsation masses 
are smaller than the evolutionary ones, but the amount of such a 
discrepancy has not been firmly established. Among the
most recent papers, we recall Bono et al. (2001, hereafter B01)
and Beaulieu et al. (2001, hereafter BBK), who studied
Cepheids in the Galaxy and in the Magellanic Clouds, respectively.

By relying on a sample of 31 variables with accurate
radii, distances and photometric parameters, B01 used theoretical
$PMR$ relations neglecting the width in temperature of the
instability strip in order to determine the pulsation mass $M_p$.
From the comparison with
evolutionary masses $M_{e,can}$ inferred by canonical (i.e., no
mass-loss, no convective core overshooting) evolutionary tracks,
they show that the ratio between $M_p$ and $M_{e,can}$ varies from
0.8 to 1, with a feeble evidence for an average discrepancy of the
order of $\sim$ 13\% for short-period Cepheids and $\sim$ 10\% for
long-period ones (see also Gieren 1989). A similar comparison was 
also performed by BBK, who investigated the huge OGLE database of 
Magellanic Cepheids (Udalski et al. 1999), using alternative choices 
for distance and reddening correction. On the basis of linear period
relations and evolutionary tracks, either canonical or with a mild
convective core overshooting, they concluded that all evolutionary
computations predict masses which are systematically larger for a fixed
luminosity, especially toward the longest periods. In this context,
let us also quote Bono et al. (2002) and Keller \& Wood (2002), who
studied LMC bump Cepheids and found that the Cepheids are $\sim$ 15\%
less massive (or $\sim$ 20\% more luminous) for their luminosity (or
mass) predicted by canonical (no overshooting) evolutionary
models. Finally, Brocato et al. (2004) investigated a selected sample
of short-period Cepheids in the LMC cluster NGC 1866 and showed that,
under reasonable assumptions for NGC 1866 reddening and distance
modulus, it appears difficult to escape the evidence for pulsation
masses smaller than the evolutionary ones, either using canonical or
mild convective core overshooting computations.

In this investigation, we shall take advantage of the sample of 34
Galactic Cepheids presented by Storm et al. (2004, hereafter S04) to
push forward the B01 result by using accurate $PLC$ relations from
updated nonlinear pulsating models, together with evolutionary
relations which account for the difference between ``static'' and
``mean" magnitudes of the pulsating stars. Actually, previous
theoretical studies for classical Cepheids (Caputo et al. 1999, Paper
IV; Caputo et al. 2000, Paper V), RR Lyrae stars (Bono et al. 1995;
Marconi et al. 2003), and anomalous Cepheids (Marconi et al. 2004)
disclosed that the discrepancy between the mean magnitude, i.e. the
time average along the pulsation cycle, and the static magnitude (the
value the variable would have in case it were a static star) is not
negligible, and increases together with the pulsation amplitude.

We present in \S 2 the pulsation models which have been used to
predict suitable analytical relations connecting the period to the
pulsator mass, mean magnitude, and color. In Section 3, the
evolutionary constraints are discussed, while \S 4 deals with
mass estimates of the observed sample of Galactic Cepheids. These
results are discussed in \S 5 taking also into account the
uncertainties due to Cepheid chemical composition, absolute distance,
and reddening. The conclusions of this investigation are briefly
outlined in \S 6.

\section{Pulsational constraints}

During the last few years, we provided theoretical predictions for
classical Cepheids as based on a wide grid of nonlinear, nonlocal, and
time-dependent convective pulsational models. The first series of
computations (Bono et al. 1999b, Paper II) includes the pulsational
properties (e.g., period and light curve) of stellar structures,  
covering a wide range of effective temperatures, stellar masses 
ranging from 5 to 11$M_{\odot}$, and a solar-like chemical composition
(Z=0.02, $Y$=0.28). For each mass, the luminosity level was fixed
according to the mass-luminosity ($ML$) relation predicted by
canonical evolutionary tracks by Castellani et al. (1992, hereafter
CCS). This theoretical framework also provides the boundaries of the
instability strip. In subsequent papers, we presented similar results, 
but for different masses, luminosities, and helium contents. That set
of models has been further implemented with new computations for the
present investigation. The assumptions on the input physics and
computing procedures have already been presented (see Bono et
al. 1999a, Paper I; Bono et al. 2000a, Paper III; Bono et al. 2000c,
Paper VI), and will not be discussed here.

The complete set of available fundamental models with Z=0.02 is listed
in Table 1\footnote{The Table 1 is only available in the on-line edition 
of the manuscript.}. For each given mass, several luminosity levels are
explored, thus covering current uncertainties on canonical $ML$
relations (Castellani et al. 1992; Bono et al. 2000b: hereafter B0), as
well as accounting for the occurrence of ``overluminous'' stellar
structures as produced by convective core overshooting and/or mass
loss. The Period-Luminosity distribution of all the $Z$=0.02
fundamental pulsators is shown in Fig. 1, where solid points display
the models computed adopting the B0 canonical $ML$ relation (see
Section 3).

Following the procedure discussed in our previous works, the
bolometric light curve of the pulsating models was transformed into
the observational bands $BVRIJK$ by means of model atmospheres by
Castelli et al. (1997a, 1997b), and these light curves are used to
derive for each pulsator the magnitude-averaged ($M_i)$ and the
intensity-averaged $\langle M_i\rangle$ magnitudes over the pulsation
cycle. Figure 2\footnote{The figures 2,3, and 11 are only available
in the on-line edition of the manuscript.} shows the ensuing $\langle M_V\rangle$ 
and $\langle M_K\rangle$ magnitudes as a function of the period. As
expected, the intrinsic scatter of the Period-Magnitude distribution,
which for any given $ML$ relation is due to the finite width of the 
instability strip, shows a substantial decrease when passing from visual 
to near-infrared magnitudes.  Concerning the distribution of the
fundamental pulsators in the color-magnitude diagram, we show in
Fig. 3 the $\langle M_V\rangle$ magnitudes versus the $\langle
M_B\rangle -\langle M_V\rangle$ colors.

It is well known that a restatement of the Stefan's law for pulsating
variables yields that the pulsation period is uniquely defined by the
mass, the luminosity, and the effective temperature of the
variable. Once bolometric corrections and color-temperature relations
are adopted, this means that the pulsator absolute magnitude $M_i$ in
a given photometric bandpass is a function of the pulsator period, 
stellar mass, and color index [$CI$], i.e.,
 
$$\langle M_i\rangle = a+b\log P+c\log M/M_{\odot}+d[CI]\eqno(1)$$

\noindent  
As a matter of fact, a linear interpolation through all the models
listed in Table 1 gives, {\it independently of any assumption on the
$ML$ relation}, tight mass-dependent $PLC$ relations, as shown in the
top panel of Fig. 4, for $\langle M_V\rangle$ magnitudes and $\langle
M_B\rangle-\langle M_V\rangle$ colors. The entire set of $PLC$
relations is given in Table 2, where the intrinsic dispersion
$\sigma_{PLC}$ includes the variation of the helium content from
$Y$=0.25 to $Y$=0.31. By using these relations, together with measured
absolute magnitudes and intrinsic colors, the pulsation mass $M_p$ of
each individual Cepheid can be determined with the intrinsic accuracy
$\epsilon_{PLC}$(log$M_p$) given in the last column of the same
table. For a Cepheid sample located at the same distance and with the
same reddening, one can estimate the mass range covered by the
variables, and in turn, the slope of the empirical $M_pL$ relation,
independently of the distance and reddening correction.  Indeed, the
BBK analysis of LMC and SMC Cepheids, which basically adopts a
``static'' $PLC$ relation, yields mass-luminosity distributions
characterized by similar slopes and intrinsic scatters for the three
different choices of distance and reddening.

A glance at the results given in Table 2 shows that the
coefficients of the color term are not dramatically different from the
extinction-to-reddening ratios $A_V/E(B-V)$=3.30, $A_R/E(V-R)$=5.29,
$A_I/E(V-I)$=1.52, $A_J/E(V-J)$=0.33, and $A_K/E(V-K)$=0.10 provided
by optical and near-infrared reddening models (see, e.g., Caldwell \&
Coulson 1987; Dean et al. 1978; Laney \& Stobie 1993; S04). This is no
surprise: as already discussed in several papers (see, e.g., Madore
1982; Madore \& Freedman 1991; Tanvir 1999; Caputo et al. 2000), the
effect of the interstellar extinction is similar to the intrinsic
scatter, due to the finite width of the instability strip. Hence, the
adoption of the various reddening insensitive Wesenheit functions
($WBV=V-3.30(B-V)$, $WVI=I-1.52(V-I)$, etc.) significantly reduces the
dispersion of magnitudes at a given period. This is shown in the
bottom panel of Fig. 4, which deals with $\langle WBV\rangle$
functions.

Assuming once again that the $ML$ relation is a free parameter, a
linear interpolation through all the fundamental models listed in
Table 1 gives the predicted mass-dependent Period-Wesenheit ($PW$)
relations listed in Table 3. These relations can be used to estimate
the pulsation mass of Cepheids with known distance, independently of
reddening. Moreover, if the variables are at the same distance, the
mass range can be derived even if a differential reddening is
present. However, it is worth noting that a residual effect, due to
the finite width of the instability strip, is still present in the
$PW$ relations (see the discussion in Madore \& Freedman 1991) and,
consequently, the intrinsic dispersion $\sigma_{PW}$ [column (4)] is
larger than $\sigma_{PLC}$ and the pulsation mass can now be
determined with lower accuracy [$\epsilon_{PW}$ in column (5)] than in
the case of pulsation masses based on the $PLC$ relations, in
particular for $B-V$ colors.

In passing, we note that the edges of the Cepheid instability strip
depend both on the $ML$ relation and on the value of the mixing-length
parameter $l/H_p$ adopted to close the system of convective transport
and conservation equations. Consequently, these two parameters affect
the predicted Period-Luminosity ($PL$) relations, mostly in the visual
bands, and play a role in the debated question of the metallicity
correction to the Cepheid intrinsic distance modulus, $\mu_0$, derived
from the $PL$ relations calibrated on LMC Cepheids (see Paper II and
Paper V). Furthermore, Fiorentino et al.  (2002, Paper VIII) have
shown that {\it sign and amount} of the predicted correction to
LMC-based distances depend on both the helium and metal content of the
variable, mainly for Cepheids with $Z>$ 0.008. In our previous papers,
we showed that the theoretical results, based on canonical $ML$
relations and $l/H_p$=1.5, supply a viable approach for reducing the
apparent discrepancy between the Cepheid and the maser distance to the
galaxy NGC 4258 (Caputo et al. 2002). The same predictions can also
account for the empirical metallicity correction $\delta
\mu_0/\delta$log$Z\sim$+0.24 mag dex$^{-1}$ derived by Kennicutt et
al. (1998), using Cepheids in two fields of the galaxy M101, provided
that a helium-to-metal enrichment ratio $\Delta Y/\Delta Z \sim$ 3.5
is adopted. Moreover, recent high resolution, high signal-to-noise
spectra for three dozen of Galactic and Magellanic Cepheids (Mottini
et al. 2004; Romaniello et al. 2004), and absolute distances based on
the near-infrared surface brightness method (S04) support the quoted
theoretical framework.  Even though current pulsation models appear to
be validated by empirical evidence, it is worth underlining that both
$PLC$ and $PW$ relations, at variance with the $PL$ relation, are
practically unaffected by the adopted $l/H_p$ value. Moreover, the
inclusion into these relations of the mass dependence overcomes the
assumption on the $ML$ relation.

Finally, we take into account metal contents slightly different than
$Z$=0.02, as suggested by individual abundance determinations for
Galactic Cepheids (see e.g., Fry \& Carney 1997; Andriewsky et al.
2002a,b,c; Luck et al. 2003; Romaniello et al. 2004). According to
fundamental models constructed by adopting $Z$=0.03 (Paper VIII) and
$Z$=0.01 (Marconi et al. 2004, in preparation), we estimated the
metallicity effect on the predicted $PLC$ and $PW$ relations.  We find
that the corrections on the estimated mass for $BV$ and $VR$ colors
are: $\Delta$log$M_p\sim -$0.35($\pm$0.03)log($Z$/0.02) and $\sim
-$0.23($\pm$0.02)log($Z$/0.02), while for $VI$, $VJ$, and $VK$ colors
are: $\Delta$log$M_p\sim+$0.02($\pm$0.02)log($Z$/0.02). The latter
values indicate that there is no significant variation, at least in
the range $Z$=0.01 to 0.03.

\section{Evolutionary constraints}

The main evolutionary properties of central He-burning
intermediate-mass stars have been extensively discussed in several
papers (see, e.g., Girardi et al. 2000, hereafter G00; B01; Castellani
et al.  2003, and references therein), and we wish only to mention
that, for fixed chemical composition and physical assumptions, the
crossing of the Cepheid instability strip occurs with a characteristic
``blue loop'', whose luminosity is almost uniquely determined by 
the original stellar mass. 
bf The reader interested in a detailed discussion concerning the dependence
of the blue loop on input physics and physical assumptions is referred
to Chiosi, Bertelli, \& Bressan (1992), Stothers \& Chin (1994), and
Cassisi (2004), and references therein.
On this ground, the relevant literature contains several theoretical 
$ML$ relations, which are widely used to estimate the Cepheid evolutionary mass.

In this investigation, using the B0 canonical evolutionary models,
which are computed with the same physics of the pulsating models, we
adopt for the canonical $ML$ relation in the mass range
4-15$M_{\odot}$ the following relation:

$$\log (L/L_{\odot})_{can}=0.72+3.35\log M/M_{\odot}+1.36\log (Y/0.28)-
0.34\log (Z/0.02)\eqno(2)$$

\noindent 
with a standard deviation $\sigma$=0.04, which accounts for both the
blueward and the redward portion of the blue loop (2$^{nd}$ and
3$^{rd}$ crossing of the Cepheid instability strip).

The introduction of a $ML$ relation, which connects the permitted
values of mass and luminosity, gives a two-parameter description of
the pulsator luminosity. Typically, as originally suggested in the
pioneering investigations by Sandage (1958), Sandage \& Gratton
(1963), and by Sandage \& Tammann (1968), the mass-term of equation
(1), if we account for evolutionary constraints, can be removed in
order to have a $PLC$ relation. However, since our purpose is to
determine the pulsation mass, from a linear interpolation through all
the fundamental models listed in Table 1, we derive the predicted
Mass-Period-Luminosity ($MPL$) and the Mass-Color-Luminosity ($MCL$)
relations given in Table 4 and in Table 5, respectively. These
relations, which are valid for structures with $Z$=0.02,
$Y$=0.28$\pm$0.03, account for the quoted uncertainty of $\sigma$=0.04
in the B0 canonical $ML$ relation, and are based on intensity-averaged
magnitudes. Moreover, they include the effects of evolutionary $ML$
relations different from equation (2). Therefore, once the $L/L_{can}$
ratio is specified, where $L_{can}$ is given by equation (2), these
relations can be used to estimate the evolutionary mass of Cepheids at
its best with an intrinsic accuracy $\epsilon$(log$M_e)\sim$ 0.03 (see
last column in Table 4 and in Table 5).

In the end of this section, two points are worth noticing:

\begin{enumerate}

\item by using the entire set of fundamental models with metal content in the range
$Z$=0.01-0.03, we find that the evolutionary mass inferred by the
predicted $MPL$ and $MCL$ relations in Table 4 and Table 5,
respectively, varies as $\Delta$log$M_e \sim$0.01log($Z$/0.02), for
fixed $L/L_{can}$ ratio;

\item    at fixed chemical composition, any change 
in luminosity relative to the canonical value $L_{can}$ leads to a
variation in the estimated evolutionary masses. Therefore, the
occurrence of overluminous stellar structures produced by convective
core overshooting (log$L/L_{can} \sim$ 0.20 at fixed mass, see Chiosi
et al. 1993) yields evolutionary masses smaller than the canonical
values. By using $MPL$ relations, we obtain $\Delta$ log$M_e \sim
-$0.03 ($K$ magnitudes) to $\sim -$0.06 ($V$ magnitudes), while with
$MCL$ relations it is $\Delta$ log$M_e \sim -0.06$ independently of
the adopted color.

\end{enumerate}

\section{Masses of Galactic Cepheids}

The recent paper by S04 gives $BVIJK$ absolute magnitudes of 34
Cepheids in the Milky Way with solar-like metal content
([Fe/H]=0.03$\pm$0.14). This means that we can determine the pulsation
mass from the predicted $PLC$ relations (see Table 2) and the
evolutionary one from the $MPL$ or the $MCL$ relations (see Table 4
and Table 5). For the sake of the following discussion, let us first
summarize in Table 6 a global estimate of the uncertainties affecting
the mass determinations as due to $ML$ relations different from
equation (2) as well as to variations of the Cepheid intrinsic
properties (period and metal content) and observational parameters
(distance and reddening).

Starting with the $PLC$ relations, we give in Table 7 the pulsation
mass determination [column (4) and columns (6) to (8)] together with
the associated error [column (5) and column (9)] as determined by the
intrinsic uncertainty of the $PLC$ relations [column (5) in Table 2]
and the error on the Cepheid intrinsic distance modulus (see S04). As
also shown in Fig. 5, where open dots refer to the short-period
variables SU Cas and EV Sct, the various estimates are in reasonable
agreement with each other, but with some evidence of the mass value
increasing, on average, when passing from $BV$ to $VK$ colors. In this
context, it should be mentioned that the reddening values adopted by
S04 came from different sources, and that any uncertainty on this
parameter affects the $BV$-based pulsation mass estimates in the
opposite way when compared with $VI$, $VJ$, and $VK$ colors (see Table
6).  Moreover, we note that by adopting for the Galactic Cepheids the
new metallicity $Z\sim$ 0.01 recently suggested for the Sun (Asplund
et al. 2004) we find that the $BV$-based pulsation masses should be
increased by $\Delta$ log$M_p\sim$ 0.11, but marginally affects the
other estimates. Therefore, we decide to adopt for the following
discussion the average value $\langle$log$M_p\rangle$ of the
$VIJK$-based mass estimates together with a final error, which
includes both the value listed in column (9) of Table 7 and the
standard error on the mean.

We now use the predicted $MPL$ relations given in Table 4 to estimate
the canonical ($L=L_{can}$) evolutionary masses at $Z$=0.02 and
$Y$=0.28$\pm$0.03. The results are listed in Table 8 [columns (4) to
(7)] together with the average associated error [column (8)] as
determined by the intrinsic uncertainty of the $MPL$ relations and the
error on the Cepheid distance. Data plotted in Fig. 6 show that there
is now a better agreement among the various estimates, even though
the estimated mass slightly decreases when moving from visual to
near-infrared magnitudes. In particular, $K$-magnitudes would lead to
masses smaller by $\Delta$ log$M_e\sim$ 0.05 than visual magnitudes. A
glance at the data listed in Table 6 suggests that the effects of
distance and reddening on the evolutionary masses based on the
predicted $MPL$ relations are $\Delta$log$M_e(V)\sim$
0.11$\Delta\mu_0$+0.36$\Delta E(B-V)$ and $\Delta$log$M_e(K)\sim $
0.26$\Delta\mu_0$+0.08$\Delta E(B-V)$. Therefore, the results plotted
in the bottom panel of Fig. 6 might suggest that, on average, the
adopted distances should be increased by $\sim$ 0.3 mag (for fixed
reddening), or the adopted reddening decreased by $\sim -$0.2 mag (for
fixed distance), or a combination of the two effects. Alternatively, a
breakdown of the canonical $ML$ relation should be considered (see
point 2 in the previous section and Table 6), with the condition
$M_e(V)$=$M_e(K)$ requiring for each given mass a luminosity increased
by $\Delta$log$L\sim$ 0.25 with respect to the canonical level, as
predicted by convective core overshooting evolutionary models.

Concerning the evolutionary mass inferred by the $MCL$ relations, the
results are listed in Table 9 [columns (4) to (7)] together with the
average associated error [column (8)]. As shown in Fig.  7, the
agreement among the various determinations is now extremely good as a
consequence of the fact that uncertainties on the adopted $ML$
relation or on the Cepheid parameters (metal content, distance, and
reddening) affect the results by almost the same quantity,
independently of the adopted color (see Table 6).  On this ground, we
adopt for the following discussion the average
$\langle$log$M_{e,can}\rangle$ of the $MCL$-based mass estimates with
a final error which includes both the value in column (8) of Table 9
and the standard error on the mean.

\section{Discussion}

The pulsation and evolutionary masses  -$\langle$log$M_p\rangle$,
$\langle$log$M_{e,can}\rangle$- estimated in \S 4, are  
presented in Fig. 8. We find that the entire Cepheid sample shows
$M_{e,can}\ge M_p$. There are three exceptions: SU Cas and EV Sct
(open dots) and the long-period variable l Car.  Moreover, data
plotted in this figure show that the discrepancy between the pulsation
and the evolutionary mass increases when moving from high to low-mass
Cepheids.

Let us briefly discuss the two short-period variables (SU Cas, EV Sct)
with $M_p>M_{e,can}$. By using the derivatives given in Table 5, we
note that a variation in the pulsation period only affects the
pulsation mass as $\Delta\langle$log$M_p\rangle \sim
-$1.13$\Delta$log$P$. Consequently, if SU Cas and EV Sct are first
overtone (FO) pulsators and their period is fundamentalised as log$P_F
\sim$ log$P_{FO}$+0.14, then the pulsation mass decreases by
$\Delta$log$M_p\sim$ 0.16, thus leading them to follow the behavior of
the other variables. This would confirm early suggestions that SU Cas
(see, e.g., Gieren 1982;  Evans 1991; Fernie et al. 1995; 
Andrievsky et al. 2002c)
and EV Sct (Tammann et al.  2003; Groenewegen et al. 2004) might be FO
pulsators. However, these two objects need to be handled with care, 
since SU Cas is connected with a reflection nebula (van den Bergh 1966) 
and EV Sct appears to show an unusual line profile structure (Kovtyukh  
et al. 2003). 

For the remaining Cepheids which are, according to Fernie et 
al. (1995), fundamental pulsators, we
plot in the top panel of Fig. 9 the ratio $M_p/M_{e,can}$ as a
function of the pulsation period. Note that, in order to minimize the
effects of uncertainties on the adopted reddening (see Table 6), we
consider only the pulsation and evolutionary mass estimates based on
$PLC(VK)$ and $MCL(VK)$ relations, respectively. Current results
suggest that the $M_p/M_{e,can}$ ratio decreases from long-period to
short-period variables, thus supporting earlier suggestions by B01 and
Gieren (1989). Data plotted in the bottom panel of the same figure
show that the inclusion of mild convective core overshooting (i.e., by
adopting log$L/L_{can}$=0.2) does not affect the pulsation masses, but
yields systematically smaller evolutionary masses, with the 
unfortunate  
consequence of several variables showing $M_p>M_{e,over}$. By the 
way, this result allows us to drop the hypothesis of noncanonical
luminosity levels, so as to solve the mild discrepancy between 
$M_{e,can}(V)$ and $M_{e,can}(K)$ values discussed in Section 3.

In order to test the dependence of current findings on the adopted
$ML$ relation, we also adopted the $ML$ relation provided by G00 
and based on canonical evolutionary computations with
$Z$=0.019, $Y$=0.273, and stellar masses in the range 4-8$M_{\odot}$.
The top panel of Fig. 10 shows the comparison between the average
luminosity predicted by G00 for canonical central
He-burning models (solid line) and the luminosity given by equation
(2) (dashed line). The two sets of models present different slopes of
the $ML$ relation (see also Fig. 5 in BBK), in particular the G00 models appear fainter for stellar masses $<5M_{\odot}$ and
brighter for masses $>5M_{\odot}$ than predictions based on B0
computations. As a consequence, (see the bottom panel in the same
figure) the adoption of the G00 canonical models leads to a
steeper dependence of the $M_p/M_{e,can}$ ratio on the Cepheid period,
and to an increased number of variables with $M_p>M_{e,can}$. The
inclusion of mild convective core overshooting makes the situation
even worse: owing to the increased luminosity for any fixed mass, all
the $M_p/M_e$ ratios become systematically larger and almost all the
Cepheids would have evolutionary masses smaller than the pulsation
ones.

The discussion of the evolutionary models is beyond the purpose of the
present paper; however, the results presented in Fig. 9 and Fig. 10
show that the current evolutionary scenario is affected not only by
the assumptions on the efficiency of overshooting, but also by
sizable differences (e.g., the equation of state) in the canonical
models. Here, relying on the S04 distance determinations, we feel that
the canonical B0 computations offer the most palatable evolutionary
scenario for studying the relation between the Cepheid pulsation mass
and the evolutionary one, {\it under the assumption of no
mass-loss}. Therefore, the trend of the $M_p/M_{e,can}$ ratio with the
Cepheid period disclosed in the top panel of Fig. 9 should be
considered as real, unless there are significant faults with our
approach or significant errors in the Cepheid adopted distance and
reddening. In order to remove any doubt on the reliability of the
adopted procedure, we use all the pulsation models listed in Table 1
as real Cepheids, and we derive their mass from the predicted $PLC$
and $MCL$ relations. We show in Fig. 11 that the ensuing ratio
between the pulsation and the evolutionary mass is
$M_p/M_{e,can}$=1$\pm$0.05, which is a quite irrelevant uncertainty
with respect to the results in Fig. 9. On the other hand, according to
the derivative values listed in Table 6, the condition
log$M_p(VK)$=log$M_{e,can}(VK)$ for the observed Cepheids would imply
rather unrealistic corrections to the adopted distance and reddening
value as given, e.g., by $\Delta\mu_0\sim$ 0.5 mag or $\Delta
E(B-V)\sim -$0.4 mag at log$P$=0.6.

To further constrain the plausibility of current theoretical predictions
we decided to perform a comparison with the dynamical mass of S Mus. This
object is the binary Cepheid with the hottest known companion and Evans
et al. (2004) by using spectra collected with both the Hubble Space
Telescope and the Far Ultraviolet Spectroscopic Explorer estimated a
mass of $M=6.0\pm0.4 M_\odot$.This mass determination agrees quite well
with the estimates of similar binary Cepheids (B{\"o}hm-Vitense et al. 1997),
but presents a smaller uncertainty. By adopting for S Mus the following input
parameters:  P=9.6599 days, $V$=6.118 (Fernie et al. 1995),
E($B-V$)=0.23 (Evans et al. 2004), $K$=3.987 (Kimeswenger et al. 2004), 
and by using the $K$-band PL relation provided by S04, we found a true 
distance modulus of $\mu_0=9.55\pm0.15$ mag.
By using these data and the PW ($V-K$) relation (see Table 3) we find
for S Mus a pulsation mass of $M=5.6\pm0.8 M_\odot$, while by assuming
$L=L_{can}$ and the MPL ($M_K$) relation (see Table 4) we find an
evolutionary mass of $6.3\pm0.6 M_\odot$. According to
Fernie et al. (1995) the reddening of S Mus is E($B-V$)=0.15, and 
in turn, the true distance modulus becomes $\mu_0=9.58\pm0.15$. Stellar 
masses based on these values are only marginally different, i.e.
$M=5.8\pm0.8 M_\odot$ (PW), $6.3\pm0.6 M_\odot$ (MPL).
Note that the uncertainties affecting current mass estimates account for
both the error on the distance modulus and for the intrinsic dispersion
of evolutionary and pulsation relation. Pulsation and evolutionary mass
agree, within the errors with the dynamical mass. However, no firm conclusion
can be reached concerning the mass discrepancy, due to current empirical
and theoretical uncertainties. An independent mass estimate for S Mus
was recently provided by Petterson, Cottrell, \& Albrow (2004), by using
high resolution spectroscopy they found $M=6.2\pm0.2 M_\odot$.
It is noteworthy, that dynamical mass of binary Cepheids might play a
crucial role in settling the discrepancy between evolutionary and
pulsation masses, since these determinations give the actual Cepheid
masses. 

In conclusion, since the pulsation mass is the actual mass of the
Cepheids, whereas the evolutionary one is based on canonical
evolutionary models neglecting mass-loss, we are quite confident that
the estimated $M_p/M_{e,can}$ ratios plotted in the top panel of
Fig. 9 reflect a mass-loss occurring during or before the central
He-burning phase. Figure 12 shows the ensuing ratio between the
difference $\Delta M=M_{e,can}-M_p$ and the canonical evolutionary
mass as a function of $M_{e,can}$. Taken at face value, the data give
sufficiently firm evidence for a mass-loss efficiency, which decreases
where increasing the Cepheid original mass. The discrepancy ranges
from $\sim$ 20\% at 4$M_{\odot}$ to $\sim$ 0 around 13$M_{\odot}$. 
This finding might appear at variance with empirical evidence, since
current semi-empirical stellar wind parametrizations indicate that the
mass-loss rate in early and in late type stars is correlated with both
stellar luminosity and radius (Reimers 1975;  Nieuwenhuijzen \&
de Jager 1990). However, current evolutionary models predict that
central He-burning phases are significantly longer when moving from
higher to lower intermediate-mass stars. In particular, the central He 
lifetime at solar chemical composition (Y=0.27, Z=0.02) increases from 
$\sim 2$ Myr for stellar structures with $M=12 M_\odot$ to $\sim 22$ Myr 
for $M=5 M_\odot$.
Moreover and even more importantly, the blue loop of the latter structure
attains cooler effective temperatures when compared with the former one.
The hottest effective temperature reached by the two structures along
the blue loop increases from $\sim 6000$ K for $M=5 M_\odot$ to 14,000 K
for $M=12 M_\odot$ (see, e.g. Table 3 and Fig. 3 in B0). These intrinsic
properties provide a plausible explanation for the increased mass loss
efficiency among short-period Cepheids.
Finally, we notice that the peculiar result
$M_p/M_{e,can}$=1.14 for the long-period variable l Car might suggest
that this Cepheid is on its first crossing of the instability
strip. In this case, a decrease in the luminosity of $\Delta$
log$L\sim -$0.2, with respect to the 2$^{nd}$ and 3$^{rd}$ crossing
luminosity, would imply $M_p \approx M_e$.

\section{Conclusions}

The comparison between theory and observations indicates that the discrepancy
between pulsation and evolutionary mass might be due to mass-loss. However,
this finding relies, as suggested by the referee, on the accuracy of
Baade-Wesselink (BW) distance determinations. A possible luminosity dependent
error cannot be excluded. In particular, angular diameters and linear variations
present a discrepancy in the phase interval between 0.8 and 1.0 (see Fig. 2
in S04). In order to overcome this problem, it has been suggested by Sabbey
et al. (1995) that the conversion factor between radial and pulsation velocity,
the so-called $p$-factor, is not constant along the pulsation cycle as assumed
in the BW method and its variants (Barnes \& Evans 1976). However, recent
time-dependent models for $\delta$ Cep by Nardetto et al. (2004) suggest
that the time dependence of the $p$-factor is marginal. Moreover and even
more importantly, we still lack firm theoretical and empirical constraints
on the dependence of the $p$-factor on the pulsation period (Gieren et al. 1993;
Marengo et al. 2004, and references therein). 

The occurrence of mass-loss was theoretically predicted by Iben (1974)
in his seminal investigation of the evolution of intermediate-mass
stars. Indeed, the computations by this author did not exclude the
possibility that these stars loose almost one third of their original
mass during the giant phase. Moreover, as suggested by Wilson \& Bowen
(1984), stellar pulsation may play a key role in causing or at least
enhancing mass-loss. In his comprehensive review, Cox (1980) discussed
the discrepancies he found by using different approaches to obtain
Cepheid masses. In particular, the pulsation masses seemed to agree
within the error with the evolutionary ones available at that time,
but the intrinsic scatter of the ratio between the two estimates was
quite large, namely $M_p/M_e=0.97\pm0.25$ for homogeneous models and
1.07$\pm0.27$ for unhomogeneous models. Gieren (1982), on the basis of
a new analysis of different methods to derive Cepheid masses, found
smaller scatters around the above ratio, but a discrepancy between
pulsation and evolutionary masses (with the former smaller than the
latter), which increases toward longer periods.

On the observational side (see Szabados 2003 for a review and
references), evidence of mass-loss during or prior the Cepheid phase
is still a rather elusive issue. Empirical estimates based on infrared
and ultraviolet emissions and VLA observations would suggest mass-loss
rates from 10$^{-10}$ to 10$^{-7}M_{\odot}$yr$^{-1}$. It is also
questionable whether the mass-loss efficiency is independent of the
pulsation period or not. As an example, IRAS data suggest roughly
constant values, whereas IUE spectra indicate that the mass-loss rate
in $\zeta$ Gem (log$P$=1.007) is 3 times smaller than the value of l
Car (log$P$=1.551). However, together with the mass-loss rate, one
should also account for the He-burning evolutionary times, which
significantly increase when decreasing the original Cepheid mass
(i.e. from long to short-period variables).

In conclusion, current empirical estimates concerning the efficiency of 
mass-loss in classical Cepheids are limited to a few objects and probably  
affected by systematic uncertainties (Deasy 1988; Szabados 2003). Moreover, 
it is not clear whether binarity might enhance the mass-loss rate. Therefore, 
the pulsational properties appear as a robust approach to get information on
mass-loss in classical Cepheids. In this context, the pulsation masses
might also provide fundamental constraints upon future evolutionary model
computations.

\acknowledgments

It is a pleasure to thank V. Castellani for several comments and
suggestions on an early draft of this paper. We also wish to 
acknowledge an anonymous referee for his/her positive comments and 
pertinent suggestions that helped us to improve the content and the 
readability of the manuscript.  
Financial support for this study was provided by PRIN~2002,2003 within 
the framework of the projects: ``Stellar populations in the Local
Group'' (P.I.: M. Tosi) and `'Continuity and Discontinuity in the
Milky Way Formation'' (P.I.: R. Gratton). This project made use
of computational resources granted by the ``Consorzio di Ricerca del
Gran Sasso'' according to the ``Progetto 6: Calcolo Evoluto e sue
applicazioni (RSV6) - Cluster C11/B''.

\pagebreak

\clearpage

\begin{table}
\caption{Intrinsic parameters for $Z$=0.02 fundamental pulsators.}
\begin{center}
\vspace{0.5truecm}
\begin{tabular}{lccccccc}
\hline \hline
      $Y$ & $M/M_{\odot}$ & log$L/L_{\odot}$ & $ML$ &  Reference\\
\hline
0.25 & 5.00 & 3.000 & B0 & This paper \\
& 7.00 & 3.490 & " & " \\
& 9.00 & 3.860 & " & " \\
& 11.00 & 4.150 & " & " \\
0.26 & 5.00 & 3.024 & 00 & This paper \\
& 7.00 & 3.512 & " & " \\
& 9.00 & 3.878 & " & " \\
& 11.00 & 4.171 & " & " \\
0.28 & 4.00 & 2.970 &   overl. & B01 \\
& 4.50 & 2.900 & CCS & " \\
& 5.00 & 3.070 & CCS & B99b\\
& 5.00 & 3.300  & overl. &  " \\
& 6.25 & 3.420 & CCS & B01\\
& 6.50 & 3.480 & " &  "\\
& 6.75 & 3.540 & " &  "\\
& 7.00 & 3.650 & " & B99b \\
& 7.00 & 3.85  & overl. &  " \\
& 9.00 & 4.000 & CCS &  " \\
& 9.00 & 4.250  & overl. &  " \\
& 11.00 & 4.400 & CCS &  " \\
& 11.00 & 4.650 & overl. &  " \\
0.31 & 5.00 & 3.130 & B0 & F02\\
& 7.00& 3.620 & " & " \\
& 9.00& 3.980 & " & " \\
& 11.00& 4.270 & " & " \\

\hline
\end{tabular}
\end{center}
Reference: Bono et al. 1999b (B99b); Bono et al. 2000 (B0); Bono et al. 2001 (B01); 
Castellani et al 1992 (CCS); Fiorentino et al. 2002 (F02).
\end{table}

\begin{table}
\caption{Predicted mass-dependent $PLC$ relations for fundamental
pulsators with fixed metal content, $Z$=0.02, and helium abundance 
ranging from $Y$=0.25 to 0.31, based on
intensity-averaged magnitudes of the pulsators. The last two
columns give the intrinsic dispersion $\sigma_{PLC}$ of the
relation and the intrinsic uncertainty $\epsilon_{PLC}$(log$M_p$) 
on the pulsation mass inferred by these relations.}
\begin{center}
\begin{tabular}{cccccc}
\hline \hline
$a$ & $b$  & $c$ & $d$ & $\sigma_{PLC}$(mag)& $\epsilon_{PLC}$(log$M_p$)\\
\hline\\
\multicolumn{6}{c}{$\langle M_V\rangle$=$a$+$b$log$P$+$c$log$M/M_{\odot}$
+$d$[$\langle B\rangle -\langle V\rangle$]} \\
$-$1.583$\pm$0.062& $-$2.800$\pm$0.045&$-$2.103$\pm$0.099 & +2.540 $\pm$0.054&   0.062 & 0.030\\\\

\multicolumn{6}{c}{$\langle M_R\rangle$=$a$+$b$log$P$+$c$log$M/M_{\odot}$
+$d$[$\langle V\rangle -\langle R\rangle$]}\\
$-$1.903$\pm$0.042& $-$2.733$\pm$0.030&$-$2.213$\pm$0.066&+4.739 $\pm$0.081&0.042  &0.020\\\\

    \multicolumn{6}{c}{$\langle M_I\rangle$=$a$+$b$log$P$+$c$log$M/M_{\odot}$+$d$[$\langle V\rangle -\langle I\rangle$]}\\
$-$2.057$\pm$0.041& $-$2.698$\pm$0.028&$-$2.266$\pm$0.064&+2.142$\pm$0.043&0.041 &0.018\\\\

    \multicolumn{6}{c}{$\langle M_J\rangle$=$a$+$b$log$P$+$c$log$M/M_{\odot}$+$d$[$\langle V\rangle -\langle J\rangle$]}\\
$-$1.707$\pm$0.038& $-$2.680$\pm$0.026& $-$2.356$\pm$0.059&+0.707$\pm$0.022&  0.038 & 0.016\\\\

    \multicolumn{6}{c}{$\langle M_K\rangle$=$a$+$b$log$P$+$c$log$M/M_{\odot}$+$d$[$\langle V\rangle -\langle K\rangle$]}\\
$-$1.605$\pm$0.040& $-$2.626$\pm$0.026& $-$2.448$\pm$0.061&+0.231$\pm$0.016&  0.040 &0.016\\\\

\hline
\end{tabular}
\end{center}
\end{table}

\begin{table}
\caption{Predicted mass-dependent $PW$ relations for fundamental
pulsators with $Z$=0.02 and $Y$=0.25 to 0.31, based on
intensity-averaged magnitudes of the pulsators. The last two
columns give the intrinsic dispersion $\sigma_{PW}$ of the
relations and the intrinsic uncertainty $\epsilon_{PW}$(log$M_p$) 
on the pulsation mass inferred by these relations.}
\begin{center}
\begin{tabular}{ccccc}
\hline \hline
$a$ & $b$  & $c$ & $\sigma_{PW}$(mag) & $\epsilon_{PW}$(log$M_p$)\\
\hline\\
\multicolumn{5}{c}{$\langle M_V\rangle -$3.30[$\langle B\rangle -\langle V\rangle$]=$a$+$b$log$P$+$c$log$M/M_{\odot}$} \\
$-$2.234$\pm$0.091& $-$3.323$\pm$0.038&$-$1.491$\pm$0.129&0.095& 0.065\\\\

\multicolumn{5}{c}{$\langle M_R\rangle -$5.29[$\langle V\rangle -\langle R\rangle$]=$a$+$b$log$P$+$c$log$M/M_{\odot}$} \\
$-$1.708$\pm$0.066& $-$2.601$\pm$0.010&$-$2.364$\pm$0.066&0.046&0.028\\\\

    \multicolumn{5}{c}{$\langle M_I\rangle -$1.52[$\langle V\rangle -\langle I\rangle$]=$a$+$b$log$P$+$c$log$M/M_{\odot}$} \\
$-$1.532$\pm$0.060& $-$2.371$\pm$0.025&$-$2.638$\pm$0.086&0.060 & 0.025\\\\

  \multicolumn{5}{c}{$\langle M_J\rangle -$0.33[$\langle V\rangle -\langle J\rangle$]=$a$+$b$log$P$+$c$log$M/M_{\odot}$} \\
$-$1.198$\pm$0.063& $-$2.320$\pm$0.026&$-$2.751$\pm$0.089&0.063 & 0.023\\\\

    \multicolumn{5}{c}{$\langle M_K\rangle -$0.10[$\langle V\rangle -\langle K\rangle$]=$a$+$b$log$P$+$c$log$M/M_{\odot}$} \\
$-$1.372$\pm$0.059& $-$2.459$\pm$0.020&$-$2.628$\pm$0.067&  0.047 & 0.022\\\\
\hline
\end{tabular}
\end{center}
\end{table}

\begin{table}
\begin{center}
\caption{ Predicted intensity-averaged $MPL$ relations
for He-burning fundamental pulsators with $Z$=0.02 and $Y$=0.25 to
$Y$=0.31. The last column gives the intrinsic uncertainty
$\epsilon_{MPL}$(log$M_e$) on the evolutionary mass inferred by
these relations due to the above helium content variation and
to the intrinsic dispersion, $\sigma$=0.04, in the adopted $ML$
relation. Note that $L_{can}$ is the luminosity of a given mass
according to the B0 canonical evolutionary tracks (see text).}
\begin{tabular}{cccccc}
\hline \hline\\

$\langle M_i\rangle$ & $a$ &$b$ & $c$ & $d$ & $\epsilon_{MPL}$(log$M_e$)\\
\hline\\
\multicolumn{6}{c}{$\langle M_i\rangle$=$a$+$b$log$P$+$c$log$M/M_{\odot}$+$d$log($L/L_{can})$}\\\\

$\langle M_V\rangle$&+3.24$\pm$0.15&+0.64$\pm$0.12      &$-$9.22$\pm$0.31&$-$2.99$\pm$0.09&0.03 \\
$\langle M_R\rangle$&+2.36$\pm$0.12&+0.06$\pm$0.10      &$-$8.04$\pm$0.29&$-$2.49$\pm$0.08&0.03 \\
$\langle M_I\rangle$&+1.59$\pm$0.10&$-$0.40$\pm$0.08    &$-$7.07$\pm$0.26&$-$2.08$\pm$0.06&0.03 \\
$\langle M_J\rangle$&+0.40$\pm$0.06&$-$1.27$\pm$0.05    &$-$5.32$\pm$0.16&$-$1.31$\pm$0.04&0.02\\
$\langle M_K\rangle$&$$-$$0.60$\pm$0.04&$-$1.95$\pm$0.04&$-$3.90$\pm$0.11&$-$0.67$\pm$0.03&0.02\\
\hline
\end{tabular}
\end{center}
\end{table}

\begin{table}
\begin{center}
\caption{ Predicted intensity-averaged $MCL$ relations
for He-burning fundamental pulsators with $Z$=0.02 and $Y$=0.25 to
$Y$=0.31. The last column gives the intrinsic uncertainty
$\epsilon_{MCL}$(log$M_e$) on the evolutionary mass inferred by
these relations due to the above helium content variation and
to the intrinsic dispersion, $\sigma$=0.04, in the adopted $ML$
relation. Note that $L_{can}$ is the luminosity of a given mass as
based on B0 canonical evolutionary tracks (see text).}
\begin{tabular}{cccccc}
\hline \hline\\

$[CI]$ & $a$ &$b$ & $c$ & $d$ & $\epsilon_{MCL}$(log$M_e$)\\
\hline\\
\multicolumn{6}{c}{$\langle M_V\rangle$=$a$+$b[CI]$+$c$log$M/M_{\odot}$+$d$log($L/L_{can})$}\\\\

$\langle M_B\rangle -\langle M_V\rangle$
&+2.42$\pm$0.11&+0.82$\pm$0.06&$-$8.35$\pm$0.11&$-$2.66$\pm$0.08&0.03 \\
$\langle M_V\rangle -\langle M_R\rangle$
&+2.28$\pm$0.11&+1.93$\pm$0.14&$-$8.37$\pm$0.10&$-$2.65$\pm$0.08&0.03 \\
$\langle M_V\rangle -\langle M_I\rangle$
&+2.24$\pm$0.11&+1.07$\pm$0.08&$-$8.37$\pm$0.10&$-$2.64$\pm$0.08&0.02 \\
$\langle M_V\rangle -\langle M_J\rangle$
&+2.32$\pm$0.10&+0.59$\pm$0.04&$-$8.39$\pm$0.10&$-$2.64$\pm$0.08&0.02 \\
$\langle M_V\rangle -\langle M_K\rangle$
&+2.33$\pm$0.10&+0.43$\pm$0.03&$-$8.39$\pm$0.10&$-$2.63$\pm$0.07&0.02 \\

\hline
\end{tabular}
\end{center}
\end{table}

\begin{table}
\caption{Estimated effects on $M_p$ and $M_e$ determinations due to variations 
in the pulsation period, the metal content, the true distance modulus, and the 
reddening.}
\begin{center}
\vspace{0.5truecm}
\begin{tabular}{lccccc}
\hline \hline
\multicolumn{6}{c}{Pulsation mass}\\  

$\partial$log$M_p$/ &log$L/L_{can}$ &$\partial$log$P$
&log$(Z/0.02)$ &$\partial\mu_0$ &$\partial
E(B-V)$\\

  $PLC(BV)$  &   --            & $-$1.33 & $-$0.35 & +0.48 &   +0.36\\
  $PLC(VI)$  &   --            & $-$1.19 & +0.02   & +0.44 & $-$0.36\\
  $PLC(VJ)$  &   --            & $-$1.14 & +0.02   & +0.42 & $-$0.40\\
  $PLC(VK)$  &   --            & $-$1.07 & +0.02   & +0.41 & $-$0.16\\
  \hline
\multicolumn{6}{c}{Evolutionary mass}\\  
$\partial$log$M_e$/ &log$L/L_{can}$ &$\partial$log$P$
&log$(Z/0.02)$ &$\partial\mu_0$ &$\partial
E(B-V)$\\
  $MPL(V)$  &   $-$0.32        &   +0.07 & +0.01   & +0.11 &   +0.36\\
  $MPL(I)$  &   $-$0.29        & $-$0.06 & +0.01   & +0.14 &   +0.28\\
  $MPL(J)$  &   $-$0.25        & $-$0.24 & +0.01   & +0.19 &   +0.15\\
  $MPL(K)$  &   $-$0.17        & $-$0.50 & +0.01   & +0.26 &   +0.08\\\\

  $MCL(BV)$ &   $-$0.32        & --      & +0.01   & +0.12 &   +0.30\\
  $MCL(VI)$ &   $-$0.32        & --      & +0.01   & +0.12 &   +0.23\\
  $MCL(VJ)$ &   $-$0.31        & --      & +0.01   & +0.12 &   +0.22\\
  $MCL(VK)$ &   $-$0.31        & --      & +0.01   & +0.12 &   +0.24\\
  \hline

\hline

\hline
\end{tabular}
\end{center}
\end{table}

\begin{table}
\begin{center}
\caption{Pulsation masses (solar units) of Milky Way Cepheids 
derived using predicted $PLC$ relations with $Z$=0.02 and $Y$=0.25 to
0.31. The errors in the last column refer to $VI$, $VJ$, and 
$VK$-based estimates.} \vspace{0.5truecm}
\begin{tabular}{lcccccccc}
\hline \hline
name&log$P$ &$M_V$ &log$M_p$ &er &log$M_p$ &log$M_p$ &log$M_p$ &er   \\
    &       &      &$\langle BV\rangle$   &   & $\langle VI\rangle$  & $\langle VJ\rangle$  & $\langle VK\rangle$  &     \\
\hline
SU~Cas   &   0.2899  &   $-$3.140  &   0.856   &   0.045   &   0.826   &   0.847   &   0.817   &   0.034   \\
EV~Sct   &   0.4901  &   $-$3.345  &   0.749   &   0.058   &   0.861   &   0.779   &   0.737   &   0.048   \\
BF~Oph   &   0.6093  &   $-$2.750  &   0.488   &   0.034   &   0.479   &   0.542   &   0.534   &   0.022   \\
T~Vel    &   0.6665  &   $-$2.692  &   0.418   &   0.041   &   0.427   &   0.526   &   0.518   &   0.030\\
$\delta$ Cep    &   0.7297 &$-$3.431  &   0.585   &   0.037   &   0.61    &   0.655   &   0.656   &   0.025   \\
CV~Mon   &   0.7307  &   $-$3.038  &   0.417   &   0.034   &   0.615   &   0.621   &   0.611   &   0.022   \\
V~Cen    &   0.7399  &   $-$3.295  &   0.531   &   0.042   &   0.583   &   0.643   &   0.638   &   0.031   \\
BB~Sgr   &   0.8220  &   $-$3.518  &   0.672   &   0.033   &   0.698   &   0.707   &   0.696   &   0.020\\
U~Sgr    &   0.8290  &   $-$3.477  &   0.633   &   0.032   &   0.660&  0.671   &   0.653   &   0.019   \\
$\eta$ Aql  &   0.8559  &   $-$3.581  &   0.577   &   0.037   &   0.61    &   0.645   &   0.639   &   0.025   \\
S~Nor    &   0.9892  &   $-$4.101  &   0.793   &   0.034   &   0.771   &   0.840&  0.822   &   0.021   \\
XX~Cen   &   1.0395  &   $-$4.154  &   0.713   &   0.032   &   0.718   &   0.770&  0.755   &   0.020   \\
V340~Nor &   1.0526  &   $-$3.814  &   0.661   &   0.093   &   0.722   &   0.719   &   0.707   &   0.020    \\
UU~Mus   &   1.0658  &   $-$4.159  &   0.691   &   0.050&  0.722   &   0.796   &   0.778   &   0.039   \\
U~Nor    &   1.1019  &   $-$4.415  &   0.727   &   0.041   &   0.735   &   0.787   &   0.771   &   0.030\\
BN~Pup   &   1.1359  &   $-$4.513  &   0.784   &   0.038   &   0.781   &   0.817   &   0.809   &   0.027   \\
LS~Pup   &   1.1506  &   $-$4.685  &   0.860   &   0.040&  0.827   &   0.874   &   0.865   &   0.029   \\
VW~Cen   &   1.1771  &   $-$4.037  &   0.675   &   0.035   &   0.713   &   0.804   &   0.786   &   0.023   \\
X~Cyg    &   1.2145  &   $-$4.991  &   1.052   &   0.031   &   0.926   &   0.941   &   0.935   &   0.020   \\
VY~Car   &   1.2768  &   $-$4.846  &   0.959   &   0.032   &   0.897   &   0.951   &   0.928   &   0.020   \\
RY~Sco   &   1.3079  &   $-$5.060  &   0.716   &   0.034   &   0.802   &   0.811   &   0.794   &   0.022   \\
RZ~Vel   &   1.3096  &   $-$5.042  &   0.858   &   0.033   &   0.845   &   0.916   &   0.897   &   0.021   \\
WZ~Sgr   &   1.3394  &   $-$4.801  &   0.866   &   0.037   &   0.892   &   0.934   &   0.915   &   0.026   \\
WZ~Car   &   1.3620  &   $-$4.918  &   0.710   &   0.043   &   0.750&  0.831   &   0.811   &   0.033   \\
VZ~Pup   &   1.3649  &   $-$5.009  &   0.643   &   0.040&  0.665   &   0.704   &   0.703   &   0.029   \\
SW~Vel   &   1.3700  &   $-$5.019  &   0.786   &   0.032   &   0.820&  0.880&  0.866   &   0.020\\
T~Mon    &   1.4319  &   $-$5.372  &   1.066   &   0.040&  0.971   &   1.028   &   1.01    &   0.029   \\
RY~Vel   &   1.4492  &   $-$5.501  &   0.909   &   0.034   &   0.905   &   0.965   &   0.93    &   0.021   \\
AQ~Pup   &   1.4786  &   $-$5.513  &   0.944   &   0.037   &   1.004   &   0.974   &   0.961   &   0.025   \\
KN~Cen   &   1.5319  &   $-$6.328  &   1.045   &   0.037   &   0.958   &   1.072   &   1.095   &   0.025   \\
l~Car    &   1.5509  &   $-$5.821  &   1.290   &   0.034   &   1.133   &   1.165   &   1.137   &   0.021   \\
U~Car    &   1.5891  &   $-$5.617  &   0.880   &   0.034   &   0.876   &   0.929   &   0.911   &   0.021   \\
RS~Pup   &   1.6174  &   $-$6.015  &   1.123   &   0.043   &   1.134   &   1.136   &   1.107   &   0.032   \\
SV~Vul   &   1.6532  &   $-$6.752  &   1.315   &   0.035   &   1.234   &   1.166   &   1.144   &   0.023   \\

\hline
\end{tabular}
\end{center}
\end{table}

\begin{table}
\begin{center}
\caption{Evolutionary masses (solar units) of Milky Way Cepheids,
derived using predicted $MPL$ relations based on canonical
evolutionary tracks with $Z$=0.02 and $Y$=0.28$\pm$0.03.}
\vspace{0.5truecm}
\begin{tabular}{lccccccc}
\hline \hline
name   &log$P$ &$M_V$ &log$M_e(V)$ &log$M_e(I)$ &log$M_e(J)$ &log$M_e(K)$ &er \\
\hline

SU~Cas   &   0.2899  &   $-$3.140 &   0.712   &   0.723   &   0.745   &   0.761   &   0.030   \\
EV~Sct   &   0.4901  &   $-$3.345    &   0.748   &   0.759   &   0.754   &   0.739   &   0.033   \\
BF~Oph   &   0.6093  &   $-$2.750 &   0.691   &   0.670   &   0.653   &   0.612   &   0.027   \\
T~Vel    &   0.6665  &   $-$2.692    &   0.689   &   0.663   &   0.648   &   0.604   &   0.029   \\
$\delta$ Cep    &   0.7297  &   $-$3.431    &   0.774   &   0.757   &   0.743   &   0.713   &   0.028   \\
CV~Mon   &   0.7307  &   $-$3.038    &   0.731   &   0.720   &   0.703   &   0.671   &   0.027   \\
V~Cen    &   0.7399  &   $-$3.295    &   0.760   &   0.742   &   0.730   &   0.698   &   0.029   \\
BB~Sgr   &   0.8220  &   $-$3.518    &   0.789   &   0.780   &   0.768   &   0.742   &   0.027   \\
U~Sgr    &   0.8290  &   $-$3.477    &   0.785   &   0.773   &   0.756   &   0.718   &   0.027   \\
$\eta$ Aql  &   0.8559  &   $-$3.581    &   0.799   &   0.779   &   0.759   &   0.718   &   0.028   \\
S~Nor    &   0.9892  &   $-$4.101    &   0.864   &   0.855   &   0.858   &   0.842   &   0.027   \\
XX~Cen   &   1.0395  &   $-$4.154    &   0.873   &   0.857   &   0.846   &   0.812   &   0.027   \\
V340~Nor &   1.0526  &   $-$3.814    &   0.837   &   0.826   &   0.807   &   0.772   &   0.044   \\
UU~Mus   &   1.0658  &   $-$4.159    &   0.876   &   0.860   &   0.855   &   0.826   &   0.031   \\
U~Nor    &   1.1019  &   $-$4.415    &   0.906   &   0.888   &   0.875   &   0.836   &   0.029   \\
BN~Pup   &   1.1359  &   $-$4.513    &   0.919   &   0.905   &   0.892   &   0.862   &   0.028   \\
LS~Pup   &   1.1506  &   $-$4.685    &   0.939   &   0.927   &   0.921   &   0.899   &   0.029   \\
VW~Cen   &   1.1771  &   $-$4.037    &   0.870   &   0.855   &   0.854   &   0.829   &   0.027   \\
X~Cyg    &   1.2145  &   $-$4.991    &   0.976   &   0.971   &   0.966   &   0.953   &   0.027   \\
VY~Car   &   1.2768  &   $-$4.846    &   0.965   &   0.958   &   0.961   &   0.945   &   0.027   \\
RY~Sco   &   1.3079  &   $-$5.060 &   0.990   &   0.971   &   0.943   &   0.888   &   0.027   \\
RZ~Vel   &   1.3096  &   $-$5.042    &   0.988   &   0.973   &   0.969   &   0.940   &   0.027   \\
WZ~Sgr   &   1.3394  &   $-$4.801    &   0.964   &   0.957   &   0.956   &   0.939   &   0.028   \\
WZ~Car   &   1.3620  &   $-$4.918    &   0.978   &   0.955   &   0.940   &   0.893   &   0.029   \\
VZ~Pup   &   1.3649  &   $-$5.009    &   0.989   &   0.955   &   0.914   &   0.842   &   0.029   \\
SW~Vel   &   1.3700  &   $-$5.019    &   0.990   &   0.972   &   0.961   &   0.926   &   0.027   \\
T~Mon    &   1.4319  &   $-$5.372    &   1.032   &   1.026   &   1.031   &   1.019   &   0.029   \\
RY~Vel   &   1.4492  &   $-$5.501    &   1.048   &   1.033   &   1.025   &   0.985   &   0.027   \\
AQ~Pup   &   1.4786  &   $-$5.513    &   1.051   &   1.046   &   1.030   &   1.003   &   0.028   \\
KN~Cen   &   1.5319  &   $-$6.328    &   1.143   &   1.123   &   1.122   &   1.113   &   0.028   \\
l~Car    &   1.5509  &   $-$5.821    &   1.089   &   1.093   &   1.108   &   1.111   &   0.027   \\
U~Car    &   1.5891  &   $-$5.617    &   1.070   &   1.050   &   1.033   &   0.987   &   0.027   \\
RS~Pup   &   1.6174  &   $-$6.015    &   1.115   &   1.116   &   1.119   &   1.108   &   0.029   \\
SV~Vul   &   1.6532  &   $-$6.752    &   1.197   &   1.200   &   1.186   &   1.164   &   0.027   \\

\hline
\end{tabular}
\end{center}
\end{table}

\begin{table}
\begin{center}
\caption{Evolutionary masses (solar units) of Milky Way Cepheids,
derived using predicted $MCL$ relations based on canonical
evolutionary tracks with $Z$=0.02 and $Y$=0.28$\pm$0.03.}
\vspace{0.5truecm}
\begin{tabular}{lccccccc}
\hline \hline
name   &log$P$ &$M_V$ &log$M_e(BV)$ &log$M_e(VI)$ &log$M_e(VJ)$ &log$M_e(VK)$ &er \\
\hline
SU~Cas&  0.2899& $-$3.140&0.706  &0.704  &0.706  &0.703  &0.027  \\
EV~Sct   &   0.4901  &   $-$3.345    &   0.735   &   0.745   &   0.737   &   0.733   &   0.029   \\
BF~Oph   &   0.6093  &   $-$2.750    &   0.679   &   0.676   &   0.681   &   0.679   &   0.025   \\
T~Vel    &   0.6665  &   $-$2.692    &   0.675   &   0.673   &   0.681   &   0.679   &   0.026   \\
$\delta$ Cep    &   0.7297  &   $-$3.431    &   0.755   &   0.755   &   0.758   &   0.758   &   0.026   \\
CV~Mon   &   0.7307  &   $-$3.038    &   0.709   &   0.725   &   0.725   &   0.723   &   0.025   \\
V~Cen    &   0.7399  &   $-$3.295    &   0.741   &   0.743   &   0.748   &   0.747   &   0.026   \\
BB~Sgr   &   0.8220  &   $-$3.518    &   0.779   &   0.780   &   0.780   &   0.779   &   0.025   \\
U~Sgr    &   0.8290  &   $-$3.477    &   0.773   &   0.774   &   0.775   &   0.772   &   0.025   \\
$\eta$ Aql  &   0.8559  &   $-$3.581    &   0.780   &   0.781   &   0.783   &   0.782   &   0.026   \\
S~Nor    &   0.9892  &   $-$4.101    &   0.854   &   0.851   &   0.857   &   0.855   &   0.025   \\
XX~Cen   &   1.0395  &   $-$4.154    &   0.857   &   0.856   &   0.860   &   0.858   &   0.025   \\
V340~Nor &   1.0526  &   $-$3.814    &   0.827   &   0.831   &   0.830   &   0.828   &   0.035   \\
UU~Mus   &   1.0658  &   $-$4.159    &   0.859   &   0.859   &   0.866   &   0.864   &   0.027   \\
U~Nor    &   1.1019  &   $-$4.415    &   0.886   &   0.885   &   0.889   &   0.887   &   0.026   \\
BN~Pup   &   1.1359  &   $-$4.513    &   0.903   &   0.901   &   0.903   &   0.902   &   0.026   \\
LS~Pup   &   1.1506  &   $-$4.685    &   0.924   &   0.920   &   0.924   &   0.922   &   0.026   \\
VW~Cen   &   1.1771  &   $-$4.037    &   0.860   &   0.861   &   0.869   &   0.867   &   0.026   \\
X~Cyg    &   1.2145  &   $-$4.991    &   0.972   &   0.960   &   0.961   &   0.960   &   0.025   \\
VY~Car   &   1.2768  &   $-$4.846    &   0.959   &   0.953   &   0.958   &   0.955   &   0.025   \\
RY~Sco   &   1.3079  &   $-$5.060    &   0.960   &   0.964   &   0.964   &   0.961   &   0.025   \\
RZ~Vel   &   1.3096  &   $-$5.042    &   0.970   &   0.967   &   0.973   &   0.971   &   0.025   \\
WZ~Sgr   &   1.3394  &   $-$4.801    &   0.955   &   0.956   &   0.960   &   0.957   &   0.026   \\
WZ~Car   &   1.3620  &   $-$4.918    &   0.954   &   0.954   &   0.961   &   0.958   &   0.026   \\
VZ~Pup   &   1.3649  &   $-$5.009    &   0.956   &   0.954   &   0.956   &   0.954   &   0.026   \\
SW~Vel   &   1.3700  &   $-$5.019    &   0.969   &   0.970   &   0.975   &   0.972   &   0.025   \\
T~Mon    &   1.4319  &   $-$5.372    &   1.027   &   1.018   &   1.023   &   1.021   &   0.026   \\
RY~Vel   &   1.4492  &   $-$5.501    &   1.027   &   1.024   &   1.029   &   1.025   &   0.025   \\
AQ~Pup   &   1.4786  &   $-$5.513    &   1.034   &   1.038   &   1.034   &   1.032   &   0.026   \\
KN~Cen   &   1.5319  &   $-$6.328    &   1.114   &   1.103   &   1.113   &   1.115   &   0.026   \\
U~Car    &   1.5891  &   $-$5.617    &   1.049   &   1.046   &   1.050   &   1.047   &   0.025   \\
l~Car    &   1.5509  &   $-$5.821    &   1.095   &   1.082   &   1.085   &   1.082   &   0.025   \\
RS~Pup   &   1.6174  &   $-$6.015    &   1.104   &   1.104   &   1.105   &   1.101   &   0.026   \\
SV~Vul   &   1.6532  &   $-$6.752    &   1.183   &   1.176   &   1.169   &   1.166   &   0.025   \\
\hline
\end{tabular}
\end{center}
\end{table}

\clearpage

\begin{figure}
\plotone{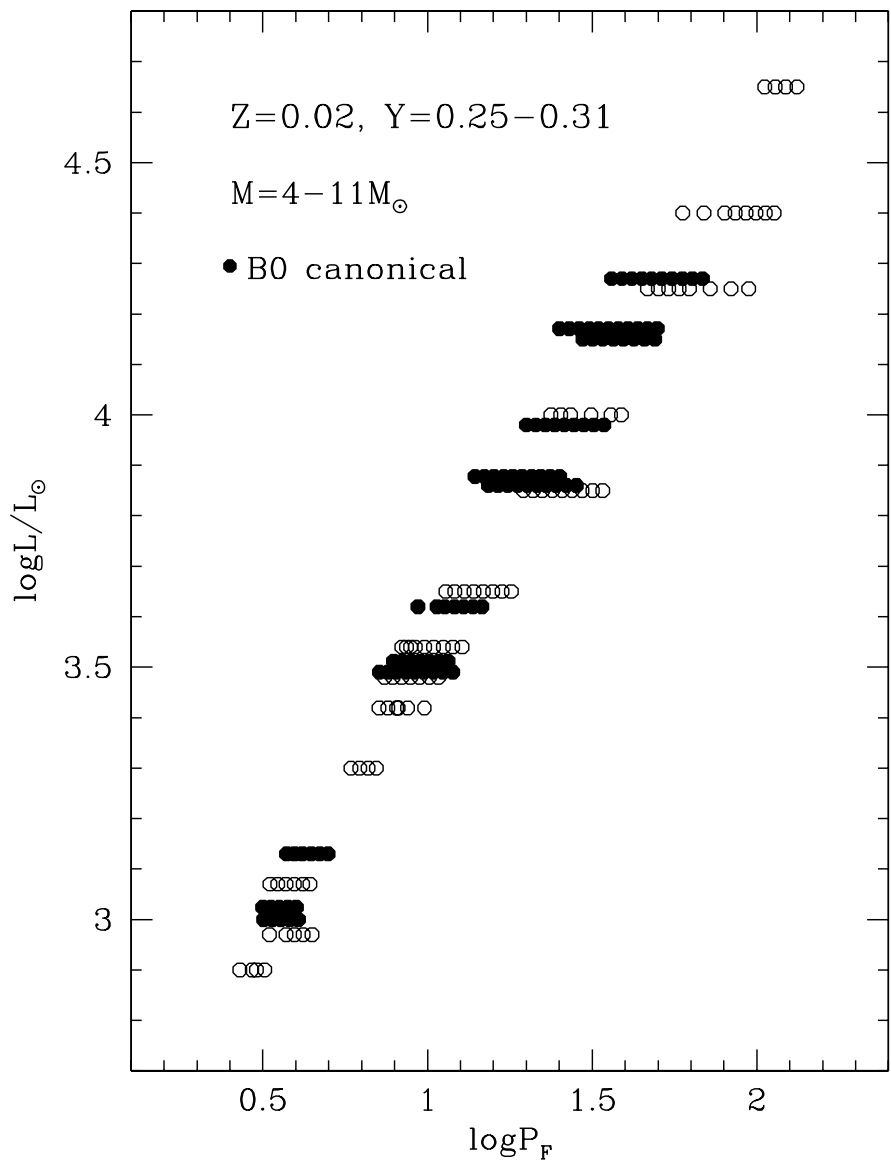}
\figcaption{Period-Luminosity distribution of fundamental pulsators with 
fixed metal content ($Z$=0.02) and helium abundance ranging from $Y$=0.25 
to 0.31. Filled dots display Cepheid models computed by adopting the 
Bono et al. (2000) canonical $ML$ relation.}
\end{figure}

\begin{figure}
\plotone{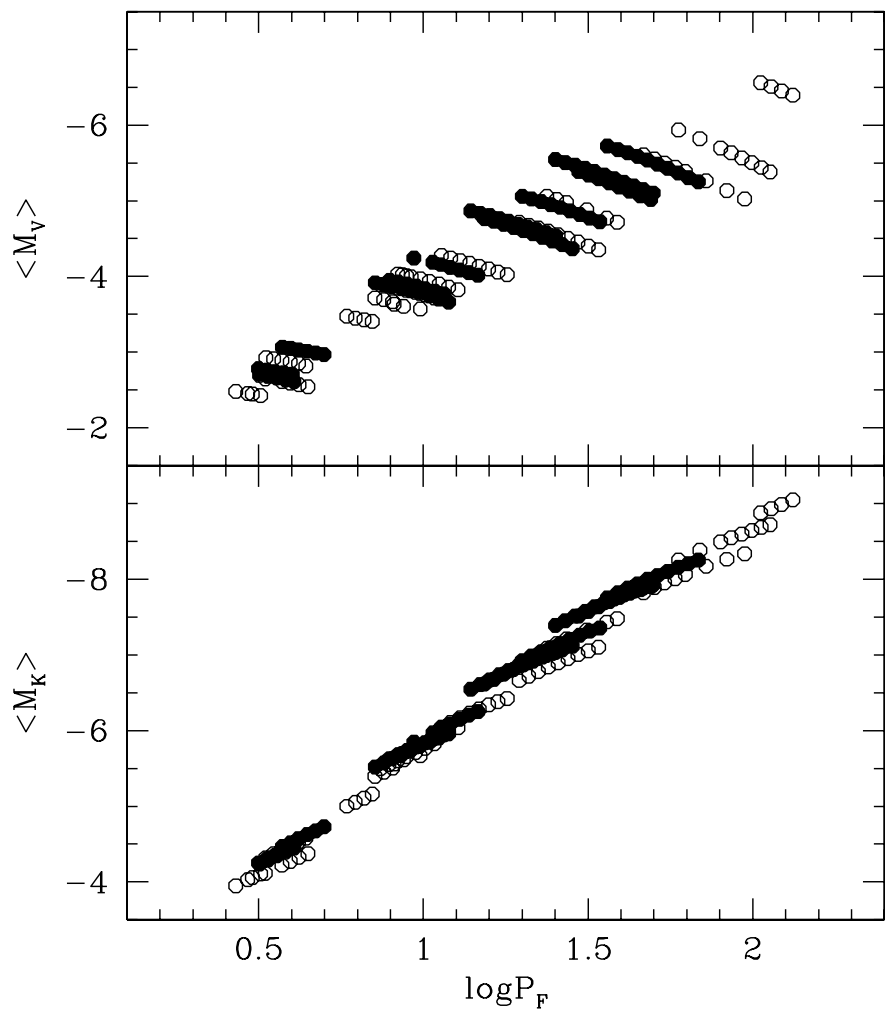}
\figcaption{Same as in Fig. 1, but for the Period-Magnitude distribution in 
V and K photometric bands.}
\end{figure}

\begin{figure}
\plotone{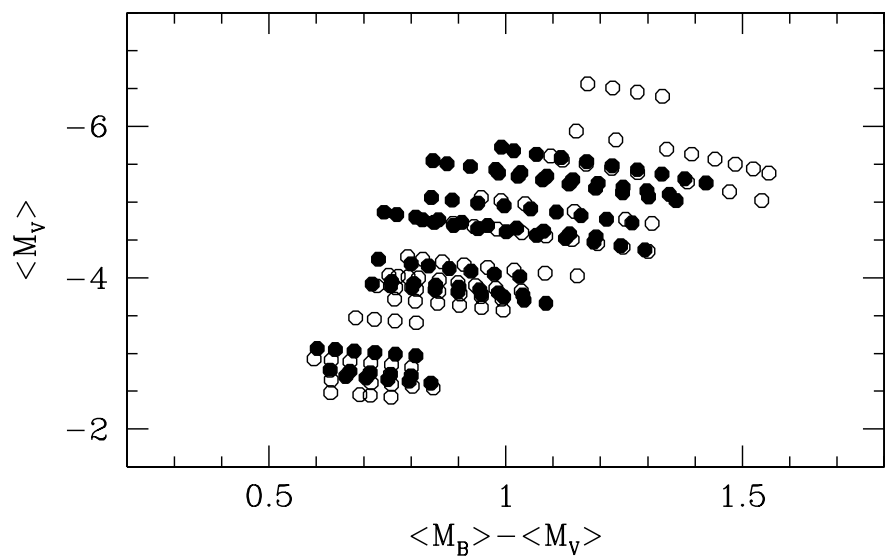}
\figcaption{Distribution of the fundamental pulsators in the Color-Magnitude 
diagram. Symbols are the same as in Figure 1.}
\end{figure}

\begin{figure}
\plotone{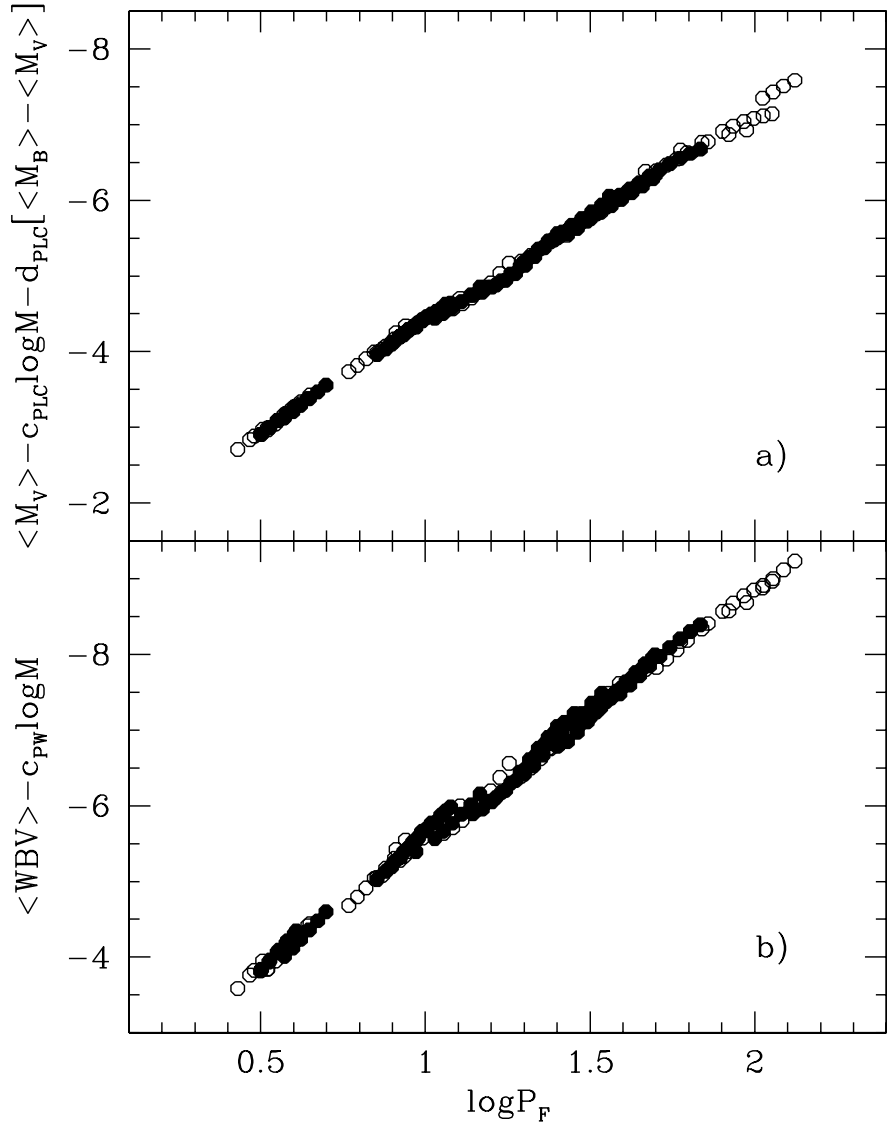}
\figcaption{Panel a) - Predicted $PLC(BV)$ relation for the fundamental
pulsators plotted in Fig. 1. Panel b) - Same as the top, but for predicted 
Period-$\langle WBV\rangle$ relation.}
\end{figure}

\begin{figure}
\plotone{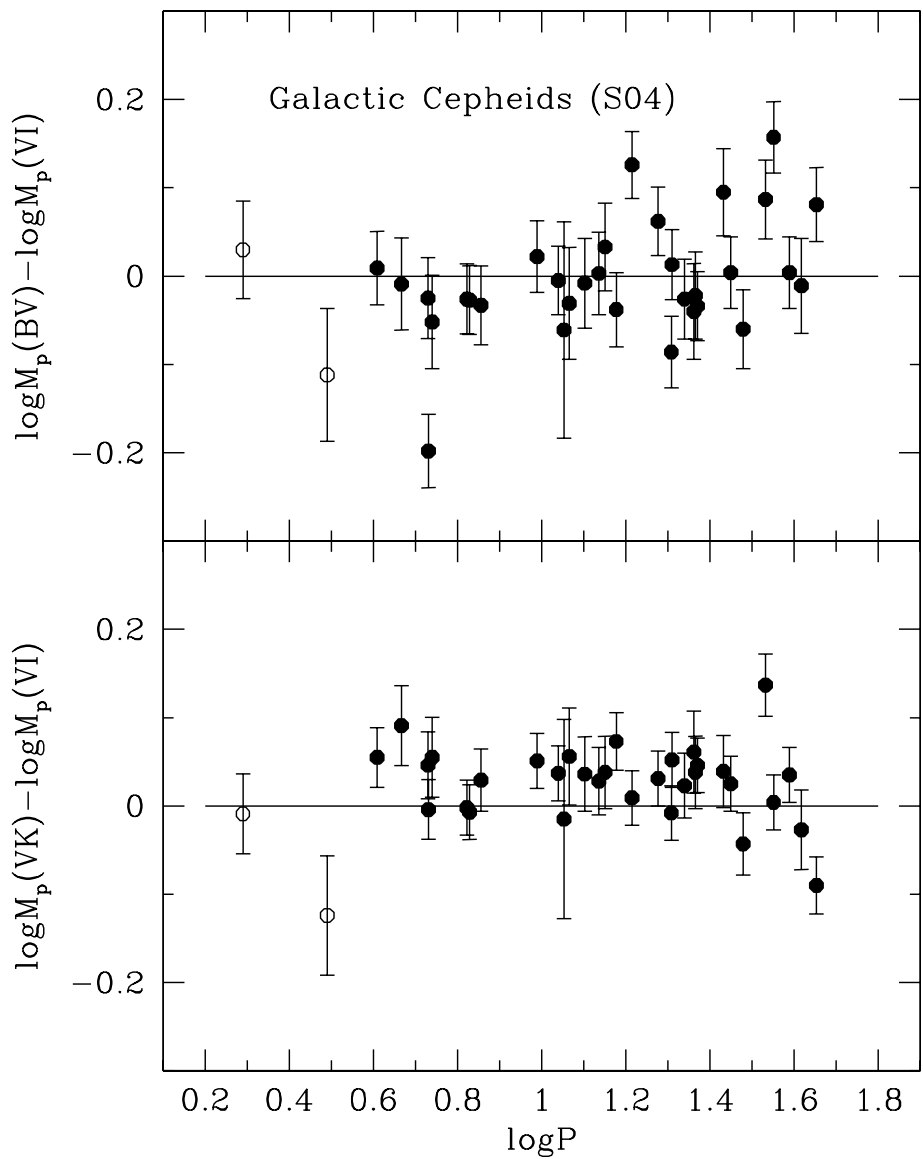}
\figcaption{Top panel - Difference in the pulsation masses of Galactic Cepheids 
estimated using optical ($B-V$, $V-I$) $PLC$ relations. Bottom panel - Same as the 
top, but for $PLC$ relations based on $V-K$ and $V-I$) colors}
\end{figure}

\begin{figure}
\plotone{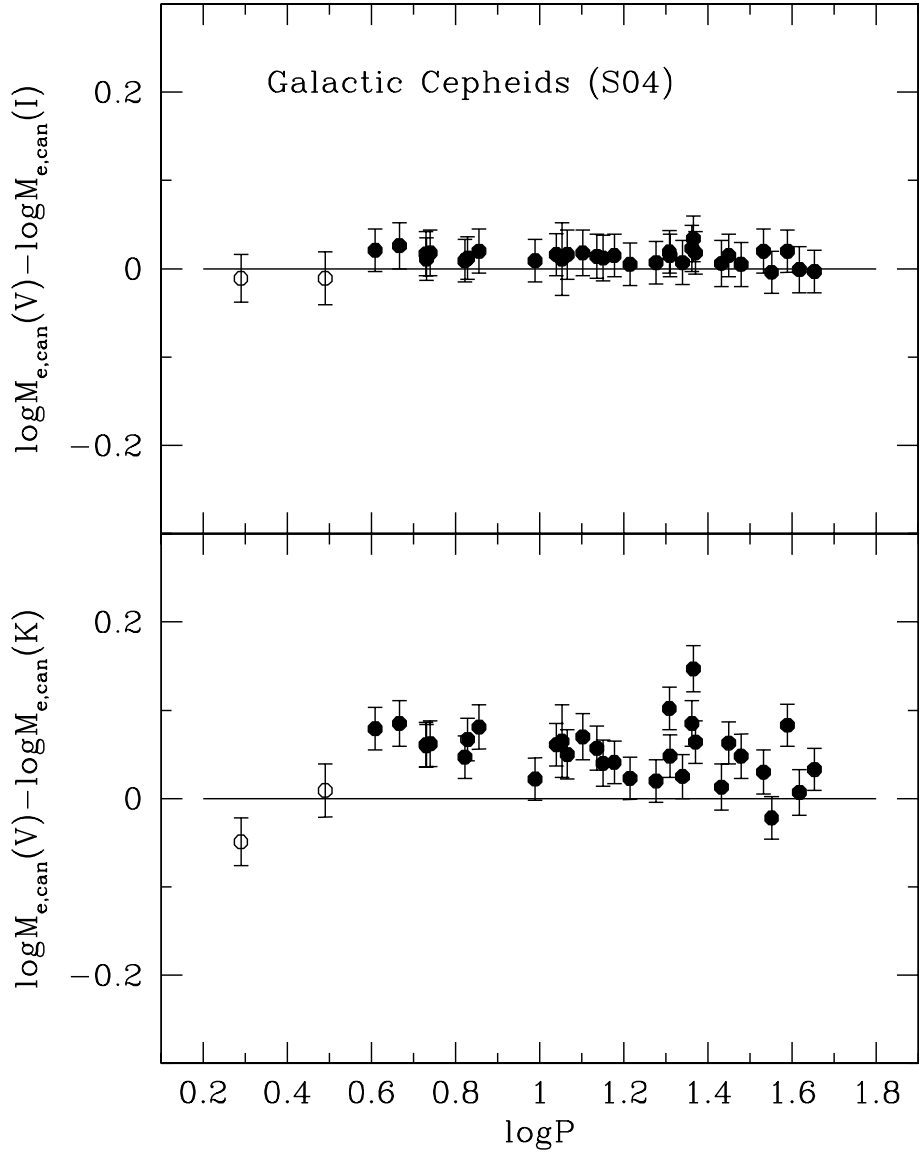}
\figcaption{Top panel - Difference in the the canonical evolutionary masses of 
Galactic Cepheids estimated using $V$ and $I$-band $MPL$ relations. Bottom panel - 
Same as the top, but for $MPL$ relations based on $V$ and $K$ magnitudes. }
\end{figure}

\begin{figure}
\plotone{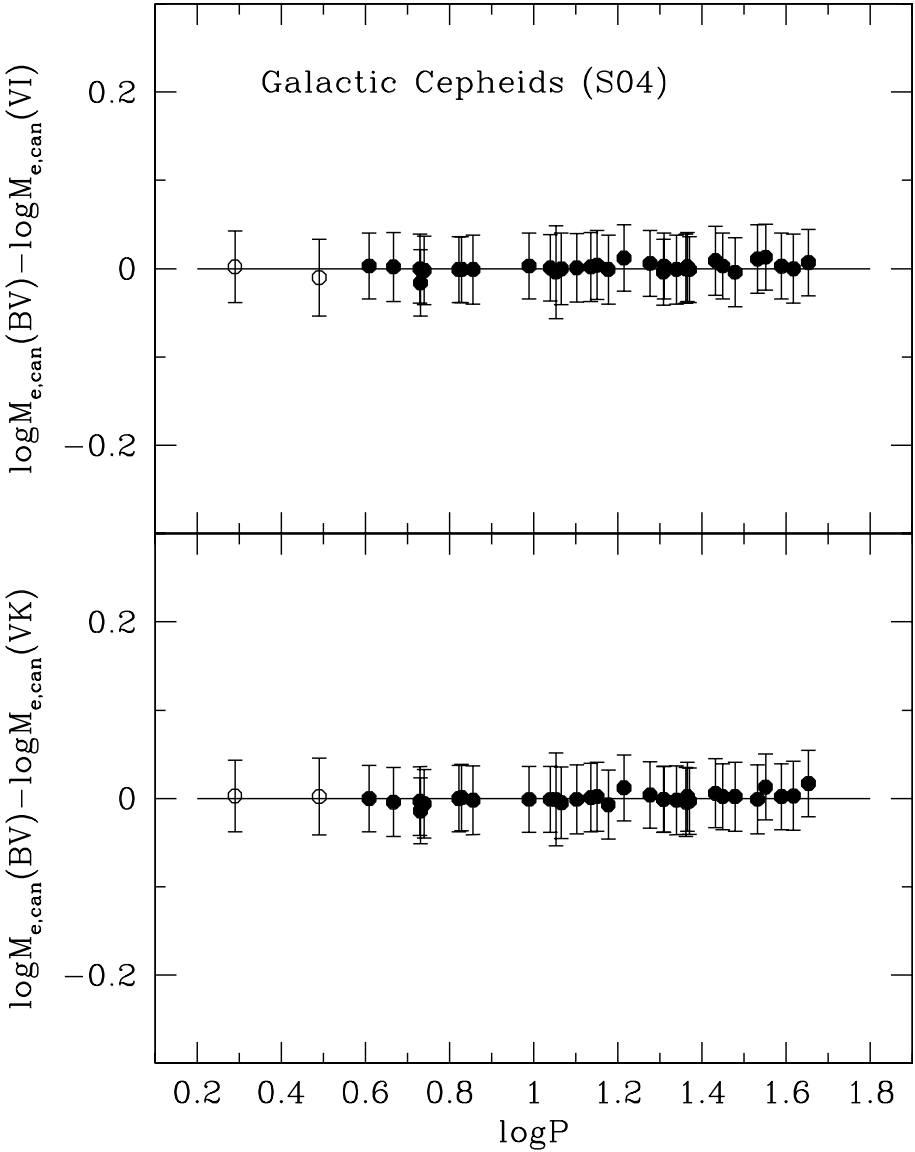}
\figcaption{Top panel - Difference in the the canonical evolutionary masses of 
Galactic Cepheids estimated using optical ($B-V$, $V-I$) $MCL$ relations. Bottom 
panel - Same as the top, but for $MCL$ relations based on $B-V$ and $V-K$ colors.}
\end{figure}

\begin{figure}
\plotone{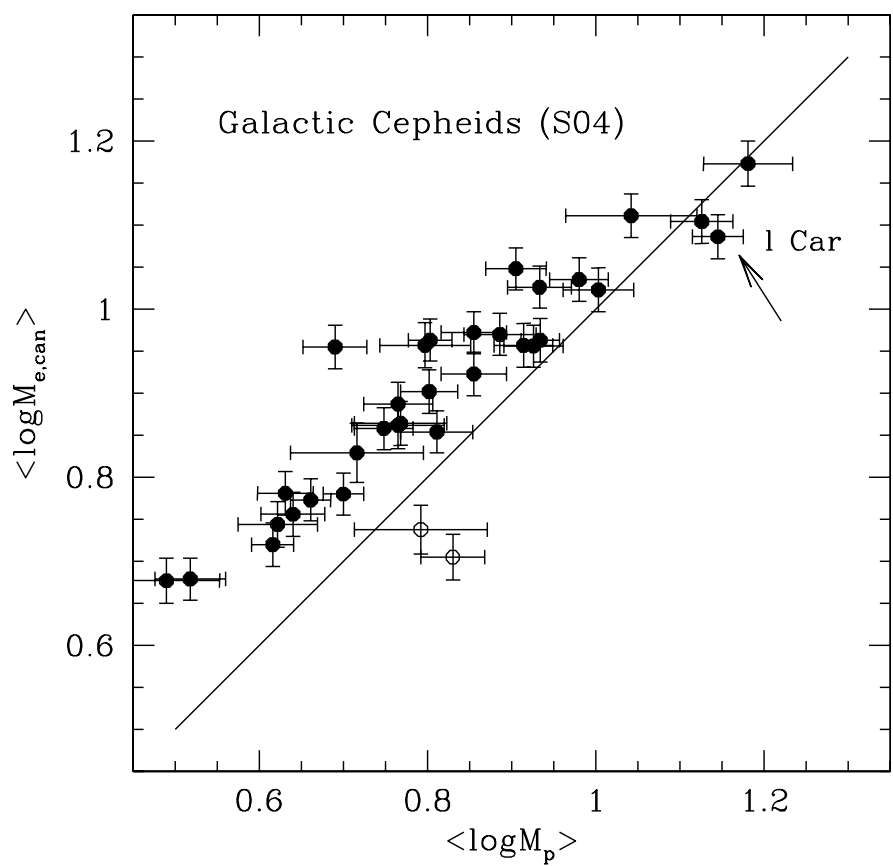}
\figcaption{Mean pulsation mass of Galactic Cepheids versus the
mean canonical evolutionary one. Open dots mark the short-period
Cepheids SU Cas and EV Sct, while the small arrow marks the
long-period variable l Car.}
\end{figure}

\begin{figure}
\plotone{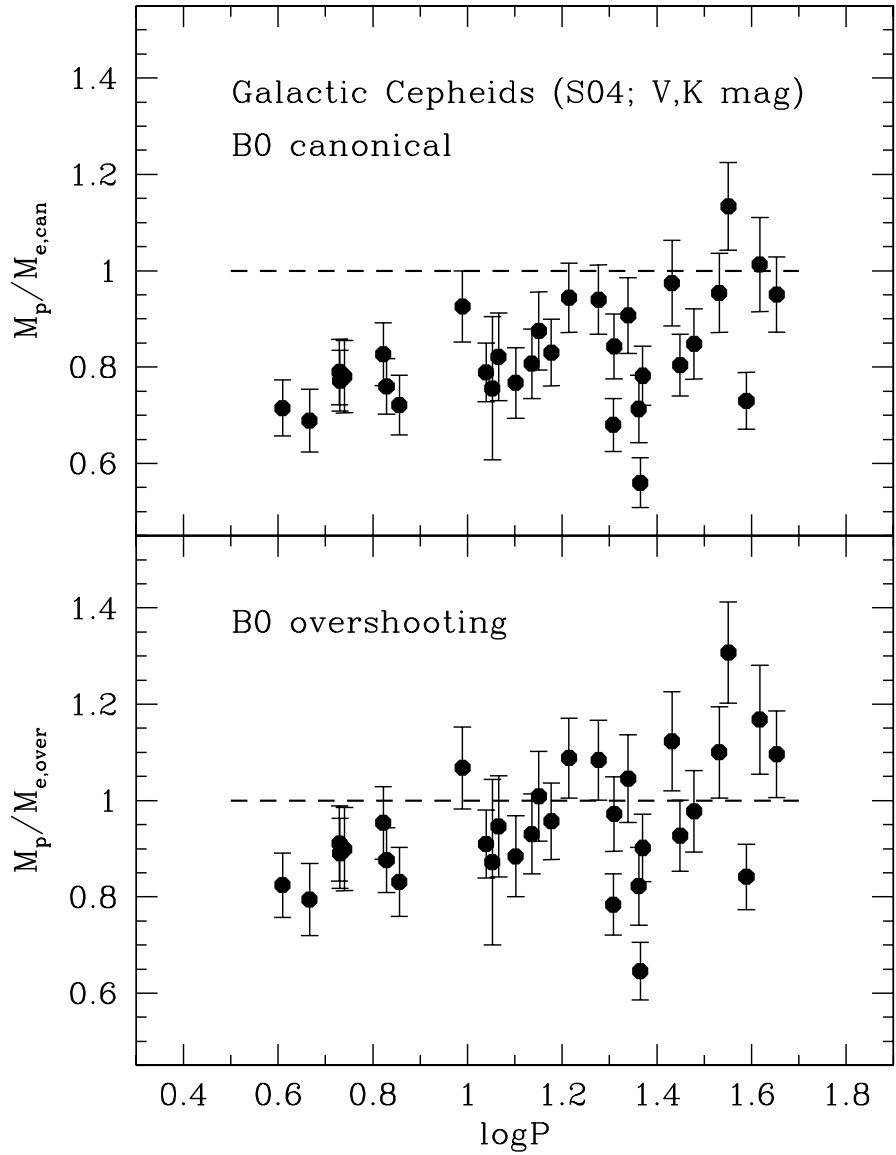}
\figcaption{Top panel: Ratio between pulsation and canonical evolutionary
mass for Galactic Cepheids as a function of period. The two
estimates are based on $V,K$ magnitudes. Bottom panel: Same as the top, 
but for evolutionary models whose luminosity was artificially increased 
to account for convective core overshooting. See text for more details.}
\end{figure}

\begin{figure}
\plotone{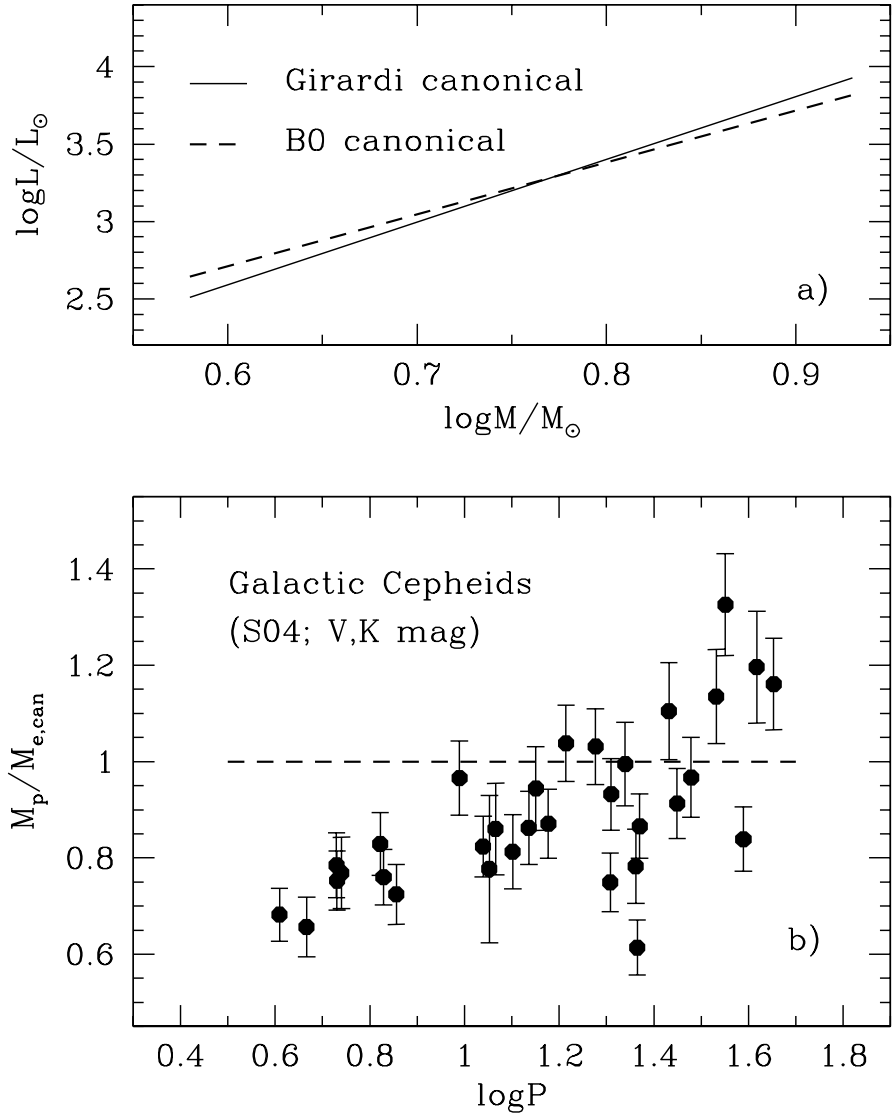}
\figcaption{Panel a) - Comparison between the canonical $ML$ relation 
adopted in this investigation (dashed line) and the $ML$ relation 
for canonical evolutionary models (solid line) provided by 
Girardi et al. (2000). Panel b) - Same as the top panel of Fig. 9, 
but based on Girardi et al. (2000) canonical evolutionary computations.}
\end{figure}

\begin{figure}
\plotone{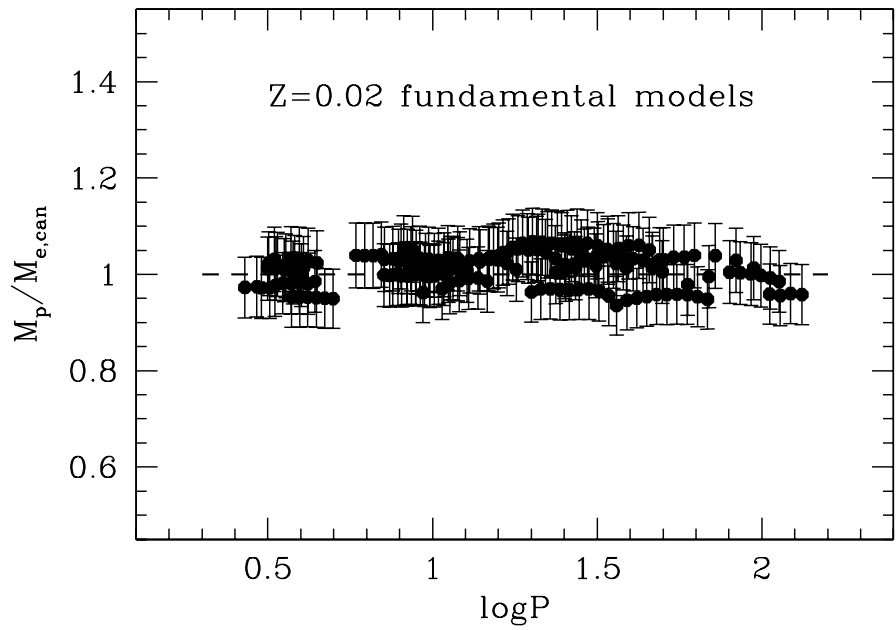}
\figcaption{Same as the top panel of Fig. 9, but for the fundamental 
pulsation models listed in Table 1.}
\end{figure}

\begin{figure}
\plotone{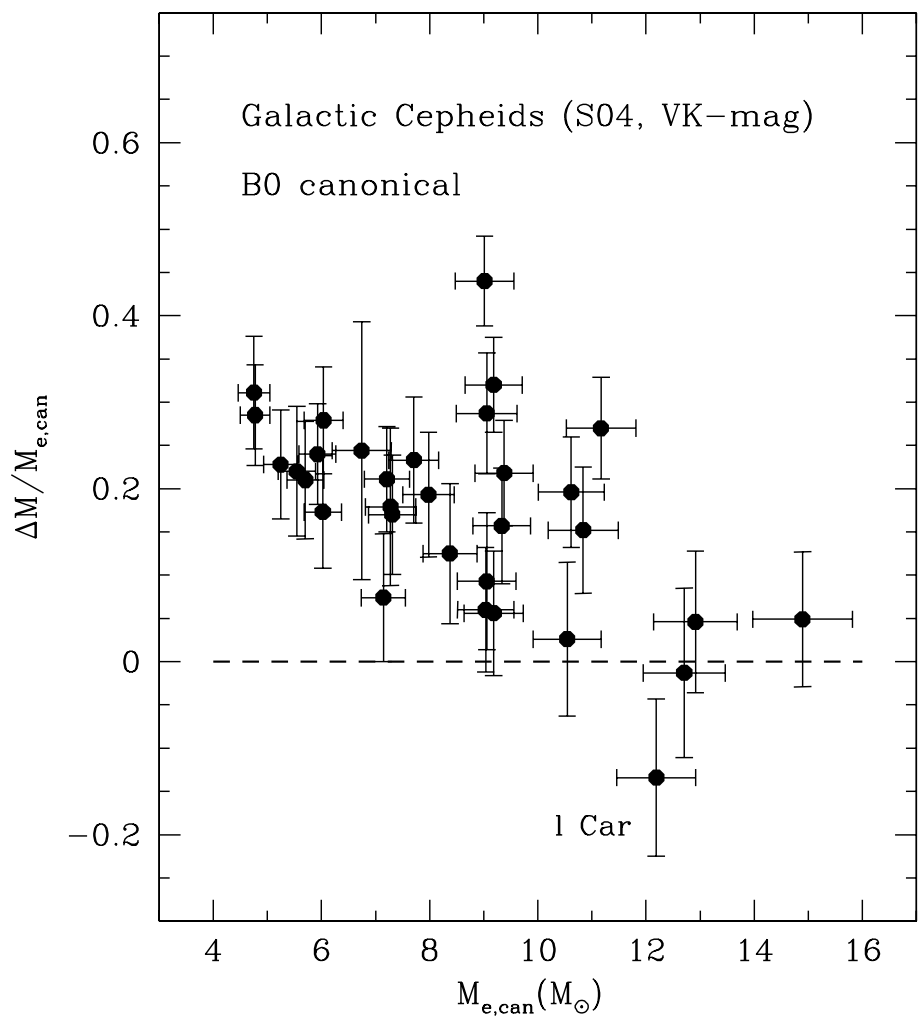}
\caption{Relative difference between pulsation ($M_p$) and canonical 
evolutionary mass ($\Delta M=M_{e,can}-M_p$) as a function of the canonical 
evolutionary mass for Galactic Cepheids.\label{f12}}
\end{figure}

\end{document}